\documentclass[8.5pt,twoside,twocolumn]{article}

\usepackage{color}

\usepackage[super,sort&compress,comma]{natbib}
\usepackage{mhchem}
\usepackage{times,mathptmx}
\usepackage{sectsty}
\usepackage{balance}
\usepackage{rotating}
\usepackage{graphicx} %eps figures can be used instead
\usepackage{lastpage}
\usepackage[format=plain,justification=raggedright,singlelinecheck=false,font=small,labelfont=bf,labelsep=space]{caption}
\usepackage{fancyhdr}
\usepackage{amsmath}

\begin{document}

\thispagestyle{plain}
\fancypagestyle{plain}{
\fancyhead[L]{\includegraphics[height=8pt]{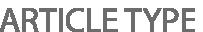}}
\fancyhead[C]{\hspace{-1cm}\includegraphics[height=20pt]{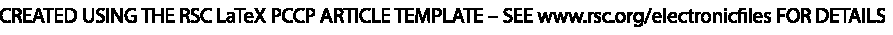}}
\fancyhead[R]{\includegraphics[height=10pt]{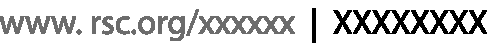}\vspace{-0.2cm}}
\renewcommand{\headrulewidth}{1pt}}
\renewcommand{\thefootnote}{\fnsymbol{footnote}}
\renewcommand\footnoterule{\vspace*{1pt}%
\hrule width 3.4in height 0.4pt \vspace*{5pt}}
\setcounter{secnumdepth}{5}

\makeatletter
\def\subsubsection{\@startsection{subsubsection}{3}{10pt}{-1.25ex plus -1ex minus -.1ex}{0ex plus 0ex}{\normalsize\bf}}
\def\paragraph{\@startsection{paragraph}{4}{10pt}{-1.25ex plus -1ex minus -.1ex}{0ex plus 0ex}{\normalsize\textit}}
\renewcommand\@biblabel[1]{#1}
\renewcommand\@makefntext[1]%
{\noindent\makebox[0pt][r]{\@thefnmark\,}#1}
\makeatother
\renewcommand{\figurename}{\small{Fig.}~}
\sectionfont{\large}
\subsectionfont{\normalsize}

\fancyfoot{}
\fancyfoot[LO,RE]{\vspace{-7pt}\includegraphics[height=9pt]{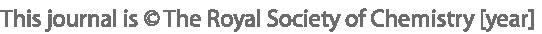}}
\fancyfoot[CO]{\vspace{-7.2pt}\hspace{12.2cm}\includegraphics{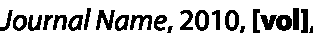}}
\fancyfoot[CE]{\vspace{-7.5pt}\hspace{-13.5cm}\includegraphics{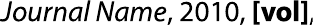}}
\fancyfoot[RO]{\footnotesize{\sffamily{1--\pageref{LastPage} ~\textbar  \hspace{2pt}\thepage}}}
\fancyfoot[LE]{\footnotesize{\sffamily{\thepage~\textbar\hspace{3.45cm} 1--\pageref{LastPage}}}}
\fancyhead{}
\renewcommand{\headrulewidth}{1pt}
\renewcommand{\footrulewidth}{1pt}
\setlength{\arrayrulewidth}{1pt}
\setlength{\columnsep}{6.5mm}
\setlength\bibsep{1pt}

\twocolumn[
  \begin{@twocolumnfalse}
\noindent\LARGE{\textbf{Elastocapillary deformations on partially-wetting substrates: rival contact-line models}}
\vspace{0.6cm}

\noindent\large{\textbf{Joshua B. Bostwick,$^{\ast}$\textit{$^{a}$} Michael Shearer,\textit{$^{b}$} and
Karen E. Daniels\textit{$^{c}$}}}\vspace{0.5cm}
%Please note that \ast indicates the corresponding author(s) but no footnote text is required.

\noindent\textit{\small{\textbf{Received Xth XXXXXXXXXX 20XX, Accepted Xth XXXXXXXXX 20XX\newline
First published on the web Xth XXXXXXXXXX 200X}}}

\noindent \textbf{\small{DOI: 10.1039/b000000x}}
\vspace{0.6cm}
%Please do not change this text.

\noindent \normalsize{A partially-wetting liquid can deform the underlying elastic substrate upon which it rests. This situation requires the development of theoretical models to describe the wetting forces imparted by the drop onto the solid substrate, particularly those at the contact-line. We construct a general solution using a displacement potential function for the elastic deformations within a finite elastic substrate associated with these wetting forces, and compare the results for several different contact-line models. Our work incorporates internal contributions to the surface stress from both liquid/solid $\Sigma_{ls}$ and solid/gas $\Sigma_{sg}$ solid surface tensions (surface stress), which results in a non-standard boundary-value problem that we solve using a dual integral equation. %Our general solution therefore encompasses all proposed contact-line models and a generalized surface energy interpretation of the wetting of soft substrates. \KED{I think this sentence is redundant with what precedes it.}
We compare our results to relevant experiments and conclude that the generalization of solid surface tension $\Sigma_{ls} \neq \Sigma_{sg}$ is an essential feature in any model of partial-wetting. The comparisons also allow us to systematically eliminate some proposed contact-line models. }
\vspace{0.5cm}
 \end{@twocolumnfalse}
  ]
\footnotetext{\textit{$^{a,\ast}$~Department of Engineering Science and Applied Mathematics, Northwestern University, Evanston, IL 60208, USA E-mail: joshua.bostwick@northwestern.edu}}
\footnotetext{\textit{$^{b}$~Department of Mathematics, North Carolina State University, Raleigh, NC 27695, USA.}}
\footnotetext{\textit{$^{c}$~Department of Physics, North Carolina State University, Raleigh, NC 27695, USA.}}
%\footnotetext{\dag~Electronic Supplementary Information (ESI) available: [details of any supplementary information available should be included here]. See DOI: 10.1039/b000000x/}
%%%%%%%%%%%%%%%%%%%%%%%%%%%%%%%%%%%%%%%%%%%%%%%%%%%%%%%%%%%%%%%%%%%%%%%%%%%%%%%%

\section{Introduction}

The deformation induced by a drop of liquid resting on a viscoelastic substrate has been studied for some time \cite{lester1961contact,shanahan1986,shanahan1987}. Describing such deformations has led to the development of the field of elastocapillarity, in which elastic stresses are coupled to surface tension (capillary forces).  Among the many biological, medical and industrial applications that involve the interaction of soft substrates with fluid interfaces \cite{Roman2010} are   enhanced condensation on soft substrates \cite{sokuler2009softer} and adhesion by liquid bridges \cite{wexler2014}. Despite much progress motivated by specific applications, a  fundamental characterization of how a liquid wets a soft viscoelastic solid remains elusive.  %\cite{marchand2012capillary},

\begin{figure}
\begin{center}
\includegraphics[width=0.24\textwidth]{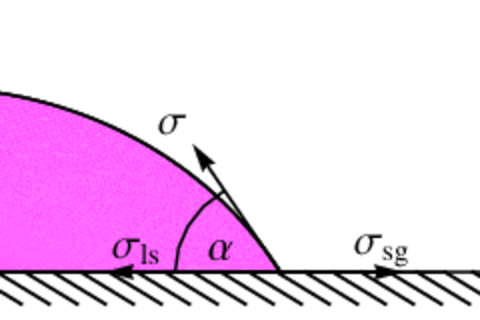}
\end{center}
\caption{\label{fig:YD} (Color online) The Young-Dupr\'{e} equation (\ref{YD}) schematically as a horizontal force balance. }
\end{figure}

\begin{figure}
\begin{center}
\begin{tabular}{c}
\includegraphics[width=0.35\textwidth]{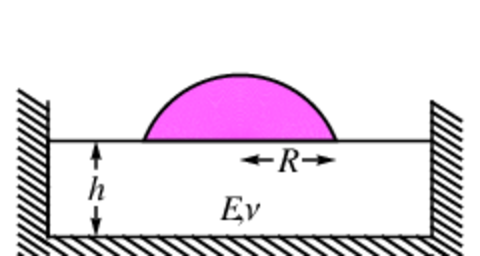} \\
\includegraphics[width=0.35\textwidth]{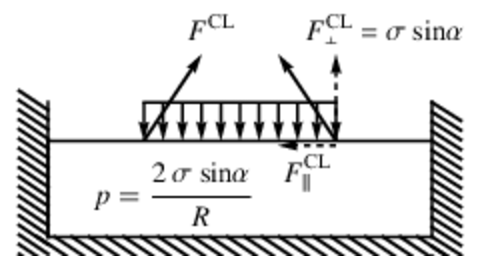}
\end{tabular}
\end{center}
\caption{\label{fig:defsketch} (Color online) Definition sketch: a liquid droplet with contact-line radius $R$ wetting an elastic substrate of height $h$, elastic modulus $E$ and Poisson ratio $\nu$ and the associated wetting forces included in the model; the capillary pressure $p$ and contact-line force $F^{\mathrm{CL}}$. }
\end{figure}

In problems coupling elasticity to capillarity, the wetting properties of the substrate strongly control the material response.
For a liquid on a hard substrate, these wetting properties are defined by the Young-Dupr\'{e} equation \cite{young,dupre},
\begin{equation}
\sigma_{sg}-\sigma_{ls}= \sigma \cos \alpha, \label{YD} \end{equation}
which relates the liquid/gas $\sigma$, liquid/solid $\sigma_{ls}$ and solid/gas $\sigma_{sg}$ surface tensions to the static contact-angle $\alpha$. Figure~\ref{fig:YD} illustrates the interpretation of the Young-Dupr\'{e} relationship as a horizontal force balance. Note that this formulation also leads to an \emph{imbalance} of forces normal to the solid substrate with magnitude $F^{\mathrm{CL}}_\perp=\sigma \sin\alpha$. The classical model of wetting of soft substrates includes this normal contact-line force applied as a point load at the contact-line, as well as the capillary pressure $p=2\sigma \sin \alpha /R$ uniformly distributed along the liquid/solid surface area, as shown in figure~\ref{fig:defsketch}. More recently, alternative models of wetting have been proposed to properly account for intrinsic surface stresses in the elastic substrate \cite{Snoeijer,weijs2013b}. For these models, thermodynamics dictates that the surface stress $\Sigma$ is related to the surface energy $\sigma$ by the Shuttleworth equation, $\Sigma_{AB}=\sigma_{AB}+\partial\sigma_{AB}/\partial \epsilon$ with $\epsilon$ the bulk strain parallel to the interface, reflecting an energetic penalty for deformation \cite{shuttleworth1950surface}. Here $A,B$ represent the phases on either side of the interface. For incompressible substrates, the surface stress $\Sigma$ is equal to the surface energy $\sigma$ and both are referred to as surface tension. Herein, we refer to $\sigma$ as the surface tension and $\Sigma$ as the solid surface tension. %In addition to the capillary pressure and normal contact-line force,
The result of the new models is to augment the classical model with a contact-line force $F^{\mathrm{CL}}_\parallel$ parallel to the solid and directed into the liquid phase.   % in addition to the capillary pressure and normal contact-line force.  %thermodynamic consistency, internal contribution to the surface stress from the solid surface tension.
%\KED{(otherwise it's not clear why a horizontal contact-line force suddenly appears)}
%Our focus is in contrasting the elastic fields induced by the different contact-line (\emph{CL}) models for the wetting forces.

%%%%%%%%%%%%%%%%%%%%%%%%%%%%%%%%%%%%%%%%%%%%%%%%%%%%%%%%%%%%%%%%%%%%%%%%%%%%%%%%
%%%%%%%%%%%%%%%%%%%%%%%%%%%%%%%%%%%%%%%%%%%%%%%%%%%%%%%%%%%%%%%%%%%%%%%%%%%%%%%%
%%%%%%%%%%%%%%%%%%%%%%%%%%%%%%%%%%%%%%%%%%%%%%%%%%%%%%%%%%%%%%%%%%%%%%%%%%%%%%%%
%%%%%%%%%%%%%%%%%%%%%%%%%%%%%%%%%%%%%%%%%%%%%%%%%%%%%%%%%%%%%%%%%%%%%%%%%%%%%%%%

In this paper we formulate a general model that describes the deformations of an elastic substrate by a partially-wetting liquid drop. The general model, formulated in terms of a displacement potential function, accommodates three rival contact-line models. By introducing both liquid/solid $\Sigma_{ls}$ and solid/gas $\Sigma_{sg}$ solid surface tensions, we generalize the work of \citet{style12} on neutrally-wetting substrates ($\alpha=90^\circ$) to partially-wetting substrates ($\alpha \neq 90^\circ$). This leads to a force boundary condition at the substrate surface that varies along the problem domain. We construct a solution to this non-standard problem by setting up a dual integral equation that results from extending the boundary condition into the complementary interval. We compare computed displacement fields from the general model to experimental results. The markedly different displacement fields predicted by the different models eliminates one model, and suggests suitable experiments to further resolve which of the others are  most plausible.

Elastocapillary phenomena generally become important when the liquid surface tension $\sigma$ and the elastic resistance of the solid substrate have similar magnitude, as measured by the elastocapillary number $\Upsilon=\sigma/EL$. Here $E$ is the elastic modulus of the substrate and $L$ is a characteristic length scale. For most liquids of interest $\sigma=10-100$~mN/m, and to adjust $\Upsilon$ it is typically easier to change $L$ or $E$. Experiments on the wrinkling of elastic sheets \cite{huang07,vella10,Davidovitch2011,schroll2013capillary} and capillary origami \cite{jung09} use small $L$. Using silicone gel \cite{jerison11,style2013universal}, gelatin \cite{spandagos12a,spandagos12b} or agar gel \cite{daniels07} as a solid substrate allows $E$ to be controllably tuned over several orders of magnitude. In systems without an intrinsic length scale, the elastocapillary length $\ell=\sigma/E$ sets the size of the elastic deformation. For reference, water ($\sigma=72$~mN/m) on a silicone gel substrate ($E=3$~kPa) yields deformations of order $\ell\sim10^{-6}$~m.

Many of the relevant experimental studies mentioned above involve neutrally-wetting ($\alpha=90^\circ$) substrates \cite{huang07,jerison11}. Studies of partial-wetting generally involve adding surfactant to the liquid to adjust the liquid/gas surface tension. \citet{schroll2013capillary} study how the wrinkling of ultra-thin elastic sheets due to a droplet is affected by the presence of a liquid bath covered in a pre-determined surfactant concentration. They derive near-threshold and far-from-threshold limits that recover experimental observations. \citet{daniels07} have shown that a droplet of surfactant-laden liquid placed on an agar gel can fracture the substrate in a starburst pattern. The number of arms in a given starburst is controlled by the surface tension contrast $\sigma_{sg}-\sigma$, an alternative measure of the degree to which a liquid partially wets a soft solid.
\citet{bostwick2013softfracture} developed a model to predict the number of arms for this situation, and have shown that the location of the contact-line, which depends upon $\alpha$ and the droplet volume $V$, is the critical parameter in wavenumber selection, in agreement with experiments. Lastly, \citet{style2013surface} study the contact mechanics of glass particles pressed into soft materials. These results show that $\alpha$ obeys a generalized Young-Dupr\'{e} equation when the indenting particle size is on the order of the elastocapillary length $\ell$.

Theory of the spreading of liquids over compliant substrates naturally relies upon an appropriate characterization of the physics of wetting, in much the same way that traditional dynamic spreading laws for liquids on rigid solids \cite{tanner79,dussan79,hocking92} build upon the static Young-Dupr\'{e} equation. For a liquid spreading on a soft viscoelastic substrate, \citet{kajiya13} show that the liquid can move continuously or with stick-slip motion depending upon the ratio of the loss to storage modulus \cite{kajiya2013wetting}. The motion of a liquid on a soft substrate experiences viscoelastic braking from the wetting ridge at the contact-line \cite{leh2012,chen2013dynamic,stapelbroek2014dynamic}.

Hence, a description of the deformation field is needed to study the dynamics of spreading. Most of the existing theoretical models are only valid for neutrally-wetting ($\alpha=90^\circ$) substrates \cite{jerison11,style12} or straight 2D contact-lines \cite{limat2012straight}. Alternative methods employ computational approaches such as density functional theory (DFT) \cite{Das2011} and molecular dynamic (MD) simulations \cite{Weijs2013} to gain a more thorough understanding of the wetting forces acting at the contact-line. The thrust of our work lies in the modeling of partially-wetting systems.
%finite elements \cite{henann2014modeling},

We begin by formulating a mathematical model (\S\ref{sec:formulation}) for the deformation of an elastic substrate due to a partially-wetting liquid droplet. The effects of partial-wetting appear in i) the contact-line force boundary conditions and ii) the solid surface tensions $\Sigma_{sg}\neq\Sigma_{ls}$. Three rival contact-line models are introduced and the governing equations are recast using a displacement potential. We then construct a general solution of the dual integral equation that results from the discontinuity in solid surface tension $\Sigma$ along the surface of the substrate. The discontinuity occurs where the interface changes from liquid/solid to solid/gas.  Our numerical results are presented in \S\ref{sec:results}, where we contrast measures of the elastic displacement field for the different models as they depend upon the model parameters. Comparisons between the predicted fields and relevant experiments allow us to systematically eliminate some proposed contact-line models. In addition, we show that the generalization of solid surface tension is an important feature for modeling wetting on soft substrates. We conclude with some remarks in \S\ref{sec:discussion} on future studies that could help resolve the issue of which model of wetting is most realistic.
%needed to reconcile the appropriate model of wetting.

%%%%%%%%%%%%%%%%%%%%%%%%%%%%%%%%%%%%%%%%%%%%%%%%%%%%%%%%%%%%%%%%%%%%%%%%%%%%%%%%

\section{Mathematical formulation \label{sec:formulation}}

A partially-wetting droplet resting on a solid substrate is held by liquid-gas surface tension $\sigma$ at its free surface. For negligible gravitational forces, the equilibrium shape is a spherical-cap with contact-line radius $R$, static contact-angle $\alpha$ and volume
\begin{equation} \frac{V}{R^3} = \frac{\pi}{3}\frac{(2-3\cos\alpha + \cos^3\alpha)}{\sin^3\alpha}.\label{SCvoldef}
\end{equation}
Note that for fixed volume drops, $R$ and $\alpha$ are not independent parameters. The linear elastic substrate has thickness $h$ and is characterized by an elastic modulus $E$ and Poisson ratio $\nu$, as shown in Figure~\ref{fig:defsketch}($a$). The liquid interacts with the solid through both the capillary pressure $p=2 \sigma \sin \alpha/R$ uniformly distributed over the liquid/solid contact area and the unbalanced contact-line force $F^{\mathrm{CL}}$ applied at the contact-line radius $R$ (c.f. Figure~\ref{fig:defsketch} ($b$)). We compute the elastic response in the substrate due to these wetting forces.
%\KED{some of the symbols in this section would be clearer with vectors (bold or arrows?). also, I wonder if $F_{\parallel}$ and $F_\perp$ would be clearer than horizontal and vertical?}
\subsection{Field equations}
We begin by introducing the axisymmetric displacement field $\boldsymbol{u}$,
\begin{equation}
\boldsymbol{u}=u_r(r,z) \boldsymbol{\hat{e}}_r +u_z(r,z) \boldsymbol{\hat{e}}_z ,\label{dispdef}\end{equation}
in cylindrical coordinates ($r,z$), which satisfies the governing elastostatic Navier equations,
\begin{equation}
\left(1-2\nu\right) \boldsymbol{\nabla}^2 \boldsymbol{u} + \boldsymbol{\nabla}\left(\boldsymbol{\nabla}\cdot \boldsymbol{u}\right) = 0. \label{navier}
\end{equation}
The strain field $\boldsymbol{\varepsilon}$ is defined as
\begin{equation}
\boldsymbol{\varepsilon} = \frac{1}{2} \left(\boldsymbol{\nabla}\boldsymbol{u} + \left(\boldsymbol{\nabla}\boldsymbol{u}\right)^t\right),\label{kinematic}
\end{equation}
while the stress field $\tau_{ij}$ for this linear elastic solid is given by
\begin{equation}
\tau_{ij} = \frac{E}{1+\nu} \left(\varepsilon_{ij} + \frac{\nu}{1-2\nu} \,\varepsilon_{kk}\right).\label{constitutive}
\end{equation}

\subsection{Boundary conditions}
We assume the elastic substrate is pinned to a rigid support at $z=0$ by enforcing a zero displacement boundary condition there,
\begin{equation}
\boldsymbol{u}(r,0)=0. \label{dispBC}
\end{equation}
On the free surface $z=h$, we specify the surface tractions
\begin{equation}\begin{split}
&\tau_{rz}(r,h)=F_r(r),\quad 0\leq r\leq \infty, \\
&\tau_{zz}(r,h) - \Sigma_{ls} \nabla^2_{\parallel}u_z (r,h) = F_z(r),\quad 0\leq r\leq R, \\
&\tau_{zz}(r,h) - \Sigma_{sg} \nabla^2_{\parallel}u_z (r,h) = F_z(r),\quad R < r\leq \infty .
\label{forceBC}\end{split}
\end{equation}
Here $\nabla^2_{\parallel}$ is the surface Laplacian and $F_z(r)$ and $F_r(r)$ are the applied vertical and horizontal forces associated with the liquid/solid interactions. These forces are model-dependent, and their particular choice will be discussed in \S\ref{s:modeldef}. As discussed by \citet{jerison11,style12}, introducing the $\Sigma$ solid surface tension (i) allows for the modeling of neutrally-wetting substrates $\Sigma_{sg}=\Sigma_{ls}$ ($\alpha=90^\circ$) and (ii) regularizes the singularity associated with applying a $\delta$-function force to the medium's surface. Here, we extend this technique to allow us to model partially-wetting substrates with $\Sigma_{sg}\neq\Sigma_{ls}$ corresponding to $\alpha\neq 90^\circ$.

\subsection{Wetting forces \label{s:modeldef}}
We now develop a model for the forces $F_z, F_r$ associated with the wetting of a liquid droplet on a soft elastic substrate. For a liquid droplet held by uniform surface tension $\sigma$, the vertical wetting forces are given by
\begin{equation} F_z(r)= \sigma \sin \alpha\left(\delta\left(r-R\right)- \frac{2}{R}\,H\left(R-r\right)\right) .\label{Vforce} \end{equation}
Here the capillary pressure $p=2 \sigma \sin \alpha /R$ (second term) is uniformly distributed over the liquid/solid surface area by the Heaviside function $H(R-r)$, whereas the unbalanced vertical contact-line force $F^{\mathrm{CL}}_z=\sigma \sin \alpha$ (first term) is applied as a point load using a delta function $\delta(r-R)$ at the contact-line $r=R$. Note the orientation of the applied forces; the capillary pressure $p$ compresses the substrate, while the contact-line force $F^{\mathrm{CL}}_z$ tends to pull the substrate upwards. Equation (\ref{Vforce}) is the standard, or classic, description of wetting of soft substrates.

More recent models of wetting have introduced an uncompensated parallel contact-line force $F_{r}(r)$, in addition to the vertical wetting forces (\ref{Vforce}) described above \cite[]{Das2011,weijs2013b}. Here we would like to construct a general solution for the models of wetting discussed below in order to contrast the resulting elastic fields. Each model for the uncompensated parallel contact-line force can be written as
\begin{equation} F_r(r) = F^{\mathrm{CL}}_r \delta(r-R) \label{Hforce}\end{equation}
with the coefficient $F^{\mathrm{CL}}_r$ for the respective model shown in Table~\ref{tab:Hforce}.
%Once again, recall that the vertical wetting forces (\ref{Vforce}) are universally accepted and common to each model.

Model I corresponds to the classic picture of wetting in which the contact line exerts no horizontal force on the substrate. In contrast, Models II and III take the same form with respect to $\alpha$, but have different dependence on the Poisson ratio $\nu$ of the substrate.  Note that for the unusual case of $\nu=0$, Model III reduces to Model II; however, ordinary materials do not typically reach this limit \cite{mott2013}. The more interesting case is that of incompressible substrates, for which $\nu=1/2$ and  Model III reduces to Model I. Many soft materials are known to be highly incompressible\cite{mark1999polymer} and modern measurement techniques\cite{pritchard2013precise} are making it possible to obtain precise values of the deviation from $1/2$. In particular, the experiments of \citet{style2013surface} report $\nu=0.495$ for the silicone gel to which we compare model results below. Even if  Model III is the correct model, the closeness to $\nu = 1/2$ would explain why
Model I has been so successful in predicting the elastic deformations on soft substrates.

%=============================================================================

\begin{table}
\centering
\begin{tabular}{cc}
%\hline
{} &  {\hspace{0.2in}$F^{\mathrm{CL}}_r$} \\
\hline
{I} &  {\hspace{0.2in}$0$}  \\
{II} &  {\hspace{0.2in}$-\sigma\left(1+\cos\alpha\right)$}  \\
{III} &  {\hspace{0.2in}$-\sigma\left(1+\cos\alpha\right)\left(\frac{1-2\nu}{1-\nu}\right)$} \\
\hline
\end{tabular}
\caption{Horizontal contact-line force $F^{\mathrm{CL}}_r$ for the classic description of wetting (I), used by \citet{jerison11,style12}, and updated models II and III, proposed by \citet{Das2011} and \citet{weijs2013b}, respectively. Here $\sigma$ is the surface tension, $\nu$ the Poisson ratio, and $\alpha$ the contact angle given by Eq.~(\ref{YD}).   }
\label{tab:Hforce}
\end{table}
%=============================================================================

\subsection{Displacement potential---Love function}

The Navier equations (\ref{navier}) are simplified by introducing the Galerkin vector $\boldsymbol{G}$ \cite[]{soutas1999elasticity}, defined such that
\begin{equation}
\boldsymbol{u} = \frac{1+\nu}{E} \left(2(1-\nu) \nabla^2\boldsymbol{G} - \nabla\left(\nabla \cdot \boldsymbol{G}\right)\right) \label{dispfunc}
\end{equation}
with
\begin{equation}
\boldsymbol{G} = \xi \left(r,z\right) \boldsymbol{\hat{e}}_z. \label{disppot}
\end{equation}
Sometimes the potential $\xi$ is referred to as the Love function from classical linear elasticity. We substitute (\ref{dispfunc}) into the coupled system of differential equations (\ref{navier}) to show that $\xi$ satisfies the biharmonic equation
\begin{equation} \nabla^4 \xi=0.\label{fieldR}
\end{equation}
The displacement (\ref{dispBC}) and traction (\ref{forceBC}) boundary conditions can similarly be written in terms of the potential function $\xi$.

\subsection{Hankel transform}

We seek solutions to (\ref{fieldR}) for the potential function using the Hankel transform pair,
\begin{subequations}\begin{eqnarray}
&& \hat{\xi} (s,z) = \int_0^{\infty}{r \xi(r,z) J_0(sr) \mathrm{d}r}, \label{hankel}\\ && \xi(r,z) = \int_0^{\infty}{s \hat{\xi}(s,z) J_0(sr) \mathrm{d}s},  \label{invhankel}
\end{eqnarray}\label{hankelpair}
\end{subequations}
where $J_0$ is the Bessel function of the first kind and $s$ is the radial wavenumber. %To recover the field $\xi$ in real space, one applies the inverse Hankel transform \begin{equation}
%\xi(r,z) = \int_0^{\infty}{s \hat{\xi}(s,z) J_0(sr) \mathrm{d}s}. \label{invhankeldef}
%\end{equation}

\subsection{Reduced equations}
We introduce the following dimensionless variables;
\begin{equation}
u \equiv \tilde{u} \frac{\sigma}{E}, ~~~
r \equiv \tilde{r}h, ~~~
 z \equiv \tilde{z}h, ~~~
 s \equiv \frac{\tilde{s}}{h}, ~~~
 R \equiv \tilde{R}h.
\label{scalings}
\end{equation}
Here lengths are scaled by the thickness of the elastic substrate $h$ and elastic deformations by the elastocapillary length $\ell \equiv\sigma/E$. Herein we drop the tildes for notational simplicity. Substituting the Hankel expansion (\ref{hankel}) into (\ref{fieldR}) gives a reduced equation for $\hat{\xi}$,
\begin{equation}
\nabla^4 \hat{\xi} = \left( \frac{d^2}{dz^2}-s^2\right)^2 \hat{\xi} = 0,\label{fieldRT}
\end{equation}
combined with the no-displacement condition on the rigid support $z=0$,
\begin{equation}
\frac{d\hat{\xi}}{dz}=0, \: (1-2\nu)\frac{d^2 \hat{\xi}}{dz^2}-2(1-\nu)s^2\hat{\xi} = 0 .\label{dispBCr2} \end{equation}
The general solution of (\ref{fieldRT},\ref{dispBCr2}) is given by
\begin{equation}
\hat{\xi} = C \left(\cosh (sz) + \frac{s z \sinh (sz)}{2(1-2\nu)}\right)+ D \left(sz \cosh (sz) - \sinh (sz)  \right), \label{fieldsol}
\end{equation}
with the constants $C,D$ to be determined from the traction boundary conditions (\ref{forceBC}). Here we note that the form of (\ref{forceBC}) is not amenable to standard analysis because the vertical boundary conditions  $\tau_{zz}$ change along the problem domain $r\in[0,\infty]$. We address this issue in the following section by constructing a solution to this non-standard problem using a dual integral formulation. Given the solution $\hat{\xi}$, we compute $\xi$ in real space by evaluating the inverse Hankel transform (\ref{invhankel}). Once the potential function $\xi$ is known, the displacement $\boldsymbol{u}$, strain $\boldsymbol{\varepsilon}$ and stress $\boldsymbol{\tau}$ fields are obtained via substitution into (\ref{dispfunc}), (\ref{kinematic}) and (\ref{constitutive}), respectively.

\subsection{Dimensionless groups}
The following dimensionless groups arise naturally from the choice of scaling (\ref{scalings}),
\begin{equation}
\Upsilon_{sg}\equiv \frac{\Sigma_{sg}}{E h}, \:\Upsilon_{ls}\equiv \frac{\Sigma_{ls}}{E h},\:\Lambda\equiv \frac{R}{h}.
\label{dimgroups}
\end{equation}
Here $\Upsilon_{sg}$ and $\Upsilon_{ls}$ are the solid/gas and liquid/solid elastocapillary numbers and $\Lambda$ is the aspect ratio or dimensionless contact-line radius. We also define the solid surface tension contrast $\Delta\Upsilon\equiv \Upsilon_{sg}-\Upsilon_{ls}$, which can be viewed as a measure of partial wetting.

\subsection{Dual integral equation}

The vertical component $\tau_{zz}$ of the traction boundary conditions (\ref{forceBC}) changes along the problem domain depending upon whether the solid substrate interacts with the liquid droplet ($r\in[0,R]$) or the passive gas ($r\in[R,\infty]$). To specify the constants $C,D$ in our general solution (\ref{fieldsol}), we recast the traction boundary conditions (\ref{forceBC}) in a form amenable to a dual integral solution,
\begin{subequations}\begin{eqnarray}&&\hspace{-0.5in} \tau_{rz}=F_r(r), \quad 0\leq r\leq \infty\label{HforceBC}\\ &&\hspace{-0.5in}
\tau_{zz}-\Sigma_{sg}\nabla^2_{\parallel} u_z-F_z (r)=\left(\Sigma_{ls}-\Sigma_{sg}\right)\nabla^2_{\parallel} u_z, \: 0\leq r\leq R \label{VforceBCinner}\\ && \hspace{-0.5in}  \tau_{zz}-\Sigma_{sg}\nabla^2_{\parallel} u_z-F_z (r)=0, \: R< r\leq \infty \label{VforceBCouter}\end{eqnarray}\end{subequations}
The vertical force balance (\ref{VforceBCinner},\ref{VforceBCouter}) is then written as
%\begin{eqnarray} && \int_0^{\infty} {A(s) J_0(sr) \mathrm{d}s}=G(r), \quad 0\leq r \leq R \nonumber \\
%&& \int_0^{\infty} {A(s) J_0(sr) \mathrm{d}s}=0, \quad R < r \leq \infty ,\label{Vforceext}\end{eqnarray}
\begin{equation}
\int_0^{\infty} {A(s) J_0(sr) \mathrm{d}s}=\begin{cases}
   G(r) &  0\leq r \leq R \\
   0       & R < r \leq \infty
  \end{cases} \label{Vforceext}
\end{equation}
with
\begin{equation} A(s)=s \left(\widehat{\tau}_{zz}+\Sigma_{sg}s^2\hat{u}_z-\hat{F_z}\right),\: G(r)= \Delta\Sigma\int_0^\infty{s^3 \hat{u}_z J_0(sr) \mathrm{d}s}, \label{dualsolcoeff}\end{equation}
%and
%\begin{equation} G(r)= \left(\Sigma_{sg}-\Sigma_{ls}\right)\int_0^\infty{s^3 \hat{u}_z J_0(sr) \mathrm{d}s}.\end{equation}
and $\Delta \Sigma\equiv \Sigma_{sg}-\Sigma_{ls}$. Equation (\ref{Vforceext}) is recognized as a dual integral equation with a standard solution \cite[]{sneddon,hoshan},
\begin{equation} A(s)=\frac{2}{\pi}\int_0^R{\cos st \int_t^R{\frac{r G(r)}{\sqrt{r^2-t^2}}\mathrm{d}r\mathrm{d}t}}. \label{dualsol}\end{equation}
Note that the solution is valid over the full domain $r\in[0,\infty]$. Substituting (\ref{fieldsol},\ref{dualsolcoeff}) into (\ref{dualsol}) yields
\begin{equation}s\left(\hat{\tau}_{zz} + \Sigma_{sg} s^2 \hat{u}_z-\hat{F_z}\right) = C A_1(s) + D A_2(s), \label{dualsol2} \end{equation}
where
\begin{equation}A_k(s)=\Delta\Sigma\frac{2}{\pi}\int_0^R{\cos st\int_0^R{\frac{r}{\sqrt{r^2-t^2}}\int_0^\infty{q^3 v_k(q) J_0(qr)\mathrm{d}q\mathrm{d}r\mathrm{d}t}}}, \label{Akdef} \end{equation}
and
\begin{subequations}\begin{eqnarray} &&\hspace{-0.5in} v_1(q)=-\frac{q^3 \sinh q}{2\left(1-2\nu\right)}, \\ &&\hspace{-0.5in} v_2(q)=q^2\frac{2(3-10\nu+8\nu^2)\sinh q-2q(1-2\nu)\cosh q}{2\left(1-2\nu\right)}.  \end{eqnarray}\label{vkdef}\end{subequations}
%\begin{equation}  v_1(q)=-\frac{q^3 \sinh q}{2\left(1-2\nu\right)}, \: v_2(q)=q^2\frac{2(3-10\nu+8\nu^2)\sinh q-2q(1-2\nu)\cosh q}{2\left(1-2\nu\right)}. \label{vkdef} \end{equation}
Equation (\ref{dualsol2}) and the Hankel-transformed horizontal force balance (\ref{HforceBC}) are a linear system of equations for $C,D$, whose solution is given in the Appendix.

\section{Results\label{sec:results}}

Our goal is to contrast the three contact-line models and the interpretation of solid surface tension for partial wetting, by comparing theoretical displacement fields to relevant experiments. Some of these comparisons can be directly evaluated using data from the literature, while others identify tests which would help design future experiments. We compute the elastic fields by substituting the coefficients $C,D$ into (\ref{disppot}) and evaluating (\ref{invhankel}) for the displacement potential, from which the displacements $\boldsymbol{u}$, strains $\boldsymbol{\varepsilon}$ and stresses $\boldsymbol{\tau}$ are readily obtained.  These solutions provide quantitative measures of how the elastic field, for instance the vertical contact-line displacement $u_z^{\mathrm{CL}}$ (peak height), varies with the model parameters.
%%%%%%%%%%%%%%%%%%%%%%%%%%%%%%%%%%%%%%%%%%%%%%%%%%%%%%%%%%%%%%%%%%%%%%%%%%%%%%%%
\begin{figure}
\begin{center}
\includegraphics[width=0.4\textwidth]{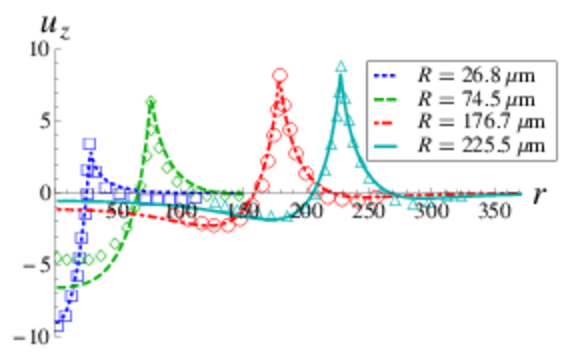}
\end{center}
\caption{\label{fig:sharpFig1comp} (Color online) Comparison with \citet[Fig. 1]{style2013universal}. Surface displacement $u_z$ on a $h=50\mu$m thick substrate against $r$ from Model I,III, as it depends upon the contact-line radius $R$ for $\nu=0.5$, $\Upsilon_{sg}=0.207$, $\Delta\Upsilon=-0.033$, $E=3$~kPa, $\sigma=46$~mN/m and $\alpha=95^\circ$. Lengths are reported in $\mu$m. Experimental results are shown with open symbols. Material properties are taken to be those reported in the experiments \cite{style2013universal}. }
\end{figure}
%%%%%%%%%%%%%%%%%%%%%%%%%%%%%%%%%%%%%%%%%%%%%%%%%%%%%%%%%%%%%%%%%%%%%%%%%%%%%%%%

We begin by comparing our model to the experimental results of \citet{style2013universal}, who use confocal microscopy to measure surface displacements on silicone gels from partially-wetting droplets. Their focus is in how the displacement fields vary with two length scales, the contact-line radius $R$ and substrate height $h$. Figure~\ref{fig:sharpFig1comp} shows how the vertical surface displacement $u_z(r,h)$ changes across the substrate for Model I. Material parameters for our computations are taken directly from the reported data \cite{style2013universal}. Note that $\nu\approx1/2$ for silicone gels, which implies that contact-line Models I and III are equivalent for these experiments. We see that our model is able to adequately reproduce the experimental results over a range of contact-line radii, which is achieved experimentally by varying the droplet volume while holding the other parameters fixed. The capillary pressure $p=2\sigma \sin \alpha /R$ tends to compress the material beneath the drop and is more pronounced for smaller drops, as would be expected. For larger drops $R=225.5\mu$m (solid line type), the contact-line force dominates the elastic response and the compressive troughs on either side of the contact-line peak become nearly symmetrical, reflecting a nearly two-dimensional solution \cite[]{jerison11}.

%%%%%%%%%%%%%%%%%%%%%%%%%%%%%%%%%%%%%%%%%%%%%%%%%%%%%%%%%%%%%%%%%%%%%%%%%%%%%%%%
\begin{figure}
\begin{center}
\includegraphics[width=0.4\textwidth]{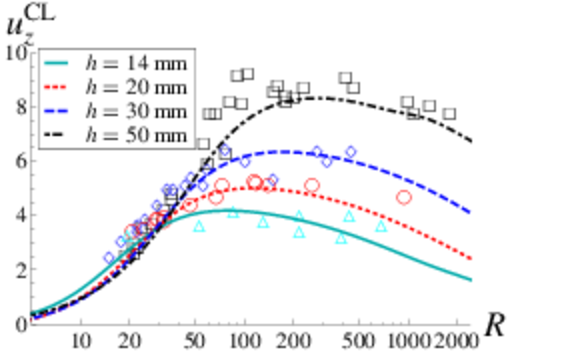}
\end{center}
\caption{\label{fig:sharpFig3comp} (Color online) Comparison with \citet[Fig. 3]{style2013universal}. Contact-line displacement $u^{\mathrm{CL}}_z$ computed from Model I,III against drop radius $R$, as it depends upon the substrate height $h$ for $\nu=0.5$, $\Upsilon_{sg}=0.207$, $\Delta\Upsilon=-0.033$, $E=3$~kPa, $\sigma=46$~mN/m and $\alpha=95^\circ$. Lengths are reported in $\mu$m. Experimental results are shown with open symbols.}
\end{figure}

\begin{figure}
\begin{center}
\begin{tabular}{ccc}
 & Models I,III & Model II \\
\begin{sideways}\hspace{.6in} \Large{$u^{\mathrm{CL}}_z$} \end{sideways} &\hspace{-0.15in}
\includegraphics[width=0.21\textwidth]{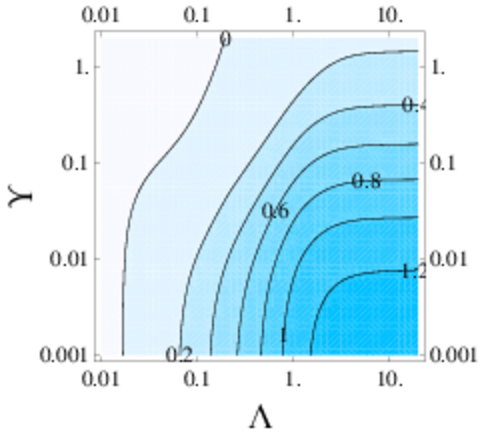} &
\includegraphics[width=0.21\textwidth]{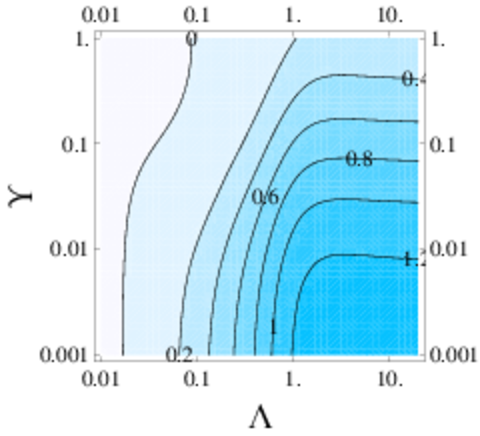} \\
\begin{sideways}\hspace{.6in} \Large{$u^{\mathrm{CL}}_r$} \end{sideways}&\hspace{-0.15in}
\includegraphics[width=0.21\textwidth]{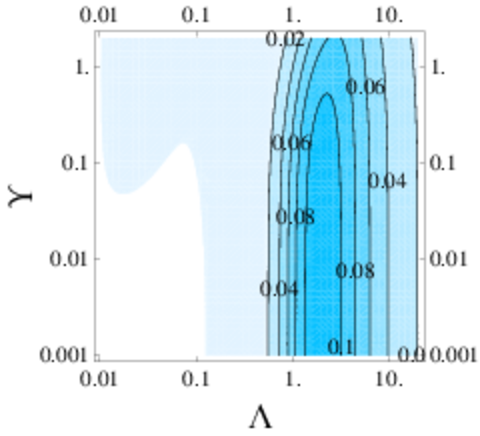} &
\includegraphics[width=0.21\textwidth]{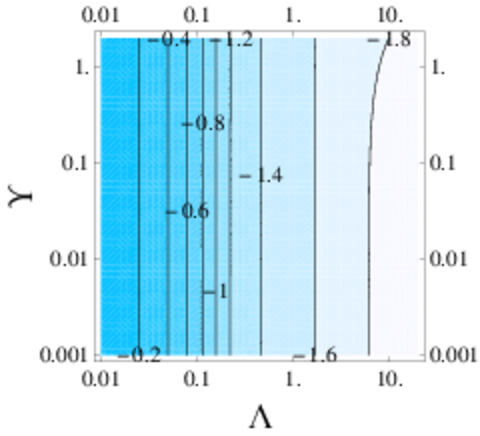} \\
\end{tabular}
\end{center}
\caption{Contact-line displacement comparing Models I,III and II by plotting the axial $u^{\mathrm{CL}}_z$ and radial $u^{\mathrm{CL}}_r$ displacement at the contact-line ($r=\Lambda$), as it depends upon the solid elastocapillary number $\Upsilon=\Upsilon_{sg}=\Upsilon_{ls}$ and the contact-line radius $\Lambda$ for $\nu=1/2$ and $\alpha=90^\circ$.  \label{fig:CLdisp90UR} }
\end{figure}

\begin{figure}[!h]
\begin{center}
\begin{tabular}{ccc}
 & Models I,III & Model II \\
\begin{sideways}\hspace{.5in} \Large{$u_z$} \end{sideways} &\hspace{-0.15in}
\includegraphics[width=0.21\textwidth]{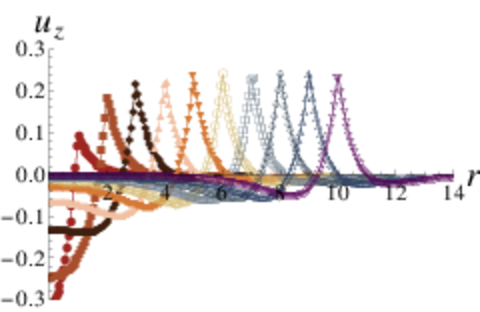} &
\includegraphics[width=0.21\textwidth]{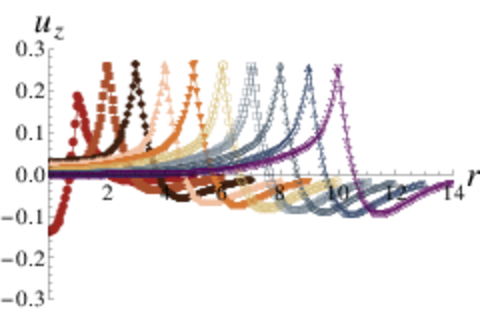} \\
\begin{sideways}\hspace{.5in} \Large{$u_r$} \end{sideways} &\hspace{-0.15in}
\includegraphics[width=0.21\textwidth]{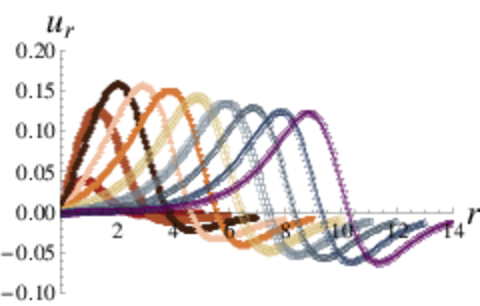} &
\includegraphics[width=0.21\textwidth]{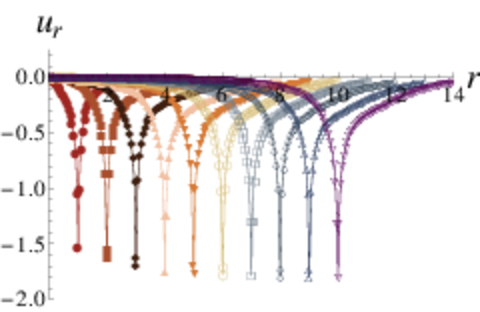}
\end{tabular}
\end{center}
\caption{\label{fig:model12comp}Comparison between Models I,III and II by plotting the surface displacements $u_z,u_r$ on an incompressible $\nu=1/2$ substrate, against $r$, as it depends upon the contact-line radius $\Lambda$, for $\Upsilon_{ls}=1, \Upsilon_{sg}=1,\alpha=90^\circ$. Note the different scales for the radial displacement.}
\end{figure}
%%%%%%%%%%%%%%%%%%%%%%%%%%%%%%%%%%%%%%%%%%%%%%%%%%%%%%%%%%%%%%%%%%%%%%%%%%%%%%%%

The peak height $u_z^{\mathrm{CL}}$ directly at the contact-line can be used as a measure of the elastic response of the underlying substrate. In Figure~\ref{fig:sharpFig3comp}, we plot the peak height for Model I as a function of the contact-line radius $R$ for various substrate heights $h$, and compare with experiments on silicone gel substrates \cite{style2013universal}. For a fixed substrate height $h$, the peak height increases with increasing contact-line radius $R$, achieves a maximum and decreases thereafter. Smaller substrate heights lead to uniformly smaller peak heights reflecting the presence of the underlying rigid support, where the zero displacement condition (\ref{dispBC}) is enforced. In contrast, thicker substrates are less affected by the underlying support since there is more material to resist the applied surface tractions, resulting in larger peak deformations. Figure~\ref{fig:sharpFig3comp} demonstrates that the non-monotonic dependence on contact-line radius also occurs in experiments. We attribute this behavior to the effects of partial wetting ($\alpha=95^\circ$) that result from a non-trivial difference between the liquid/solid and solid/gas solid surface tensions $\Delta\Upsilon\neq 0$.

For neutrally-wetting $\alpha=90^\circ$ substrates with $\Delta\Upsilon=0$, the peak height is a monotonic function of the contact-line radius $\Lambda$ for Model I, as shown in Figure~\ref{fig:CLdisp90UR}. That is, the peak height increases with the contact-line radius and then plateaus. Consequently, we can rule out Model I with $\Delta\Upsilon=0$, since it does not reproduce the experimental data. In contrast, Figure~\ref{fig:CLdisp90UR} shows that for Model II, the peak height is a non-monotonic function of the contact-line radius that is also consistent with experiments (c.f. Figure~\ref{fig:sharpFig3comp}).  We conclude that the generalization which differentiates between the $\Sigma_{ls}$ and $\Sigma_{sg}$ solid surface tensions is an essential feature of any model for elastocapillary deformations.

We proceed by contrasting contact-line Models I and II on incompressible ($\nu=1/2$) neutrally-wetting ($\alpha=90^\circ$) substrates. Additional substrate profiles for $\alpha\neq90^\circ$ are given in the Supplementary Material. Figure~\ref{fig:model12comp} plots the surface displacements $u_z,u_r$ for Models I and II, respectively. The peak heights $u_z^{\mathrm{CL}}$ are superficially similar, but the fields vary greatly away from the contact-line. Notice that beneath the drop the field is compressive for Model I and tensile for Model II. Outside the drop, the compressive dimple is much more pronounced for Model II. A more dramatic difference between Model I and II is seen in the radial surface displacement. For Model I, there is a peak on the droplet side and a trough on the gas side of the contact-line that eventually becomes symmetric as the contact-line radius increases. In contrast, the radial displacement is directed into the drop ($u_r<0$) for Model II. In addition to the qualitative differences in the radial displacement field, note the radial displacement $u_r$ scale changes by an order of magnitude between Models I and II. This observation is robust and occurs over a large range of parameters. Such a dramatic effect should clearly be visible in experiment. However, \citet[Fig. 2]{jerison11} measure the radial displacement field on incompressible silicone gel substrates showing a field more similar to that of Model I than Model II. We conclude that contact-line Model II does not accurately capture the existing experimental data and, hence, rule it out as a candidate contact-line law.

%%%%%%%%%%%%%%%%%%%%%%%%%%%%%%%%%%%%%%%%%%%%%%%%%%%%%%%%%%%%%%%%%%%%%%%%%%%%%%%%
\begin{figure}
\begin{center}
\begin{tabular}{cc}
\large{$u_z$} & \large{$u_r$}  \\
\includegraphics[width=0.23\textwidth]{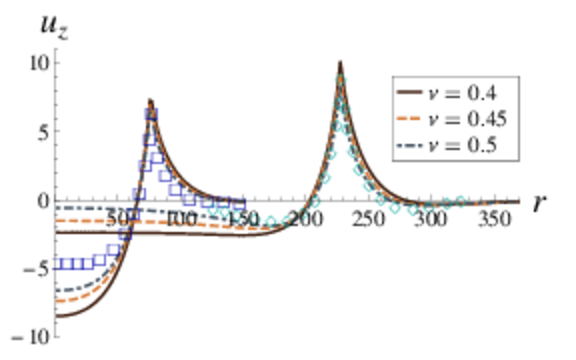} &
\includegraphics[width=0.23\textwidth]{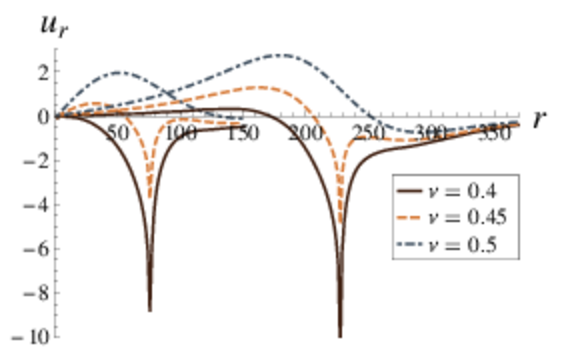}
\end{tabular}
\end{center}
\caption{\label{fig:model3comp} Compressibility effects from contact-line Model III with $R=74.5\mu$m and $R=225.5\mu$m: axial $u_z$ and radial $u_r$ displacement field in $\mu$m, as it depends upon the Poisson ratio $\nu$, for $\Upsilon_{sg}=0.207, \Delta\Upsilon=-0.033$, $E=3$kPa, $\sigma=46$mN/m and $\alpha=95^\circ$. Open symbols in sub-figure are two experiments from \citet{style12}.   }
\end{figure}

\begin{figure}
\begin{center}
\begin{tabular}{ccc}
&Model I  & Model III \\
\begin{sideways}\hspace{.65in} \Large{$u^{\mathrm{CL}}_z$} \end{sideways}&\hspace{-0.15in}
\includegraphics[width=0.21\textwidth]{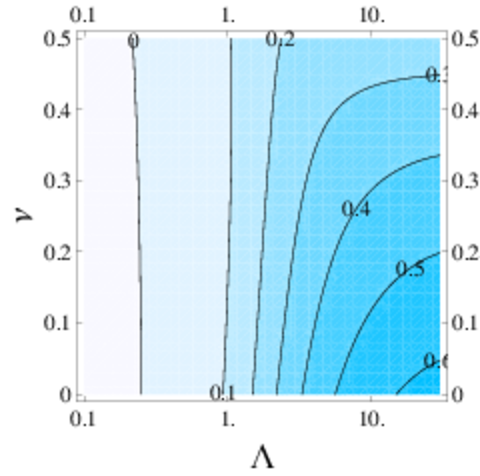} &
\includegraphics[width=0.21\textwidth]{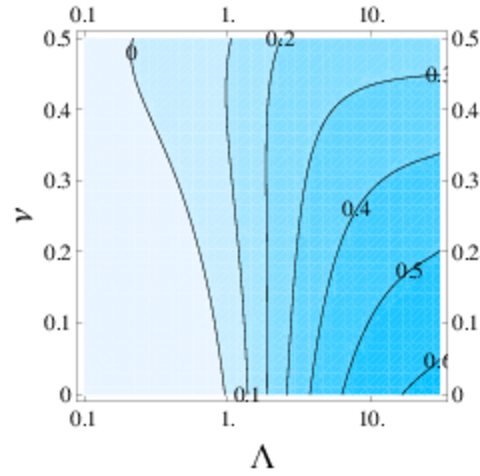} \\
\begin{sideways}\hspace{.65in} \Large{$u^{\mathrm{CL}}_r$} \end{sideways}&\hspace{-0.15in}
\includegraphics[width=0.21\textwidth]{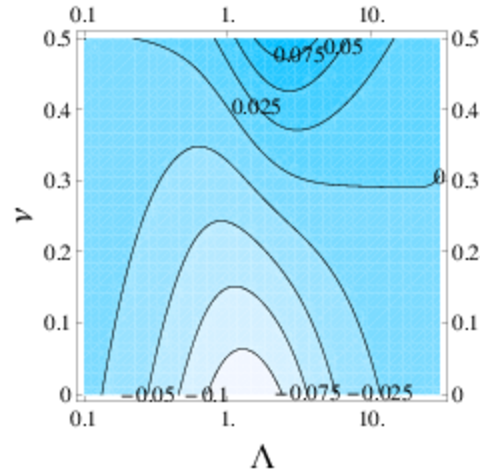} &
\includegraphics[width=0.21\textwidth]{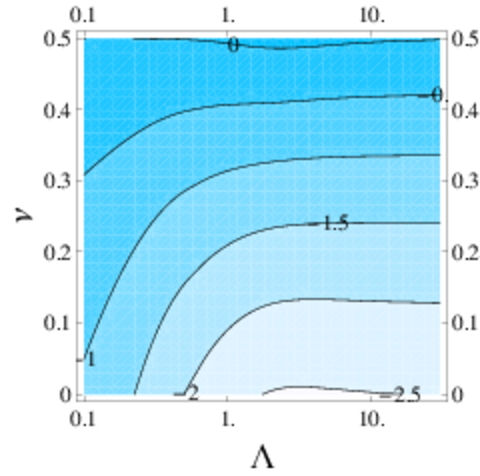}
\end{tabular}
\end{center}
\caption{Comparison of the axial $u^{\mathrm{CL}}_z$ and radial $u^{\mathrm{CL}}_r$ contact-line displacement for Models I and III, as it depends upon the Poisson ratio $\nu$ and the contact-line radius $\Lambda$ for $\Upsilon\equiv\Upsilon_{sg}=\Upsilon_{ls}=1$ and $\alpha=90^\circ$.  \label{fig:CLdispNR}}
\end{figure}
%%%%%%%%%%%%%%%%%%%%%%%%%%%%%%%%%%%%%%%%%%%%%%%%%%%%%%%%%%%%%%%%%%%%%%%%%%%%%%%%

At this point, our candidate models have been reduced to either Model I or Model III. We have demonstrated above that the generalization of solid surface tensions ($\Delta\Upsilon\neq0$) is an essential feature of any model. Recall that Model III includes a horizontal force that depends upon the Poisson ratio $\nu$, which degenerates into Model I when $\nu=1/2$. Figure~\ref{fig:model3comp} shows the displacement fields for Model III depend upon $\nu$ for nearly incompressible substrates. For the vertical displacement $u_z$, the peak height does not appreciably change with $\nu$, while the largest difference occurs near the center of the drop $r=0$. The most dramatic difference occurs for the radial displacement $u_r$, where the presence of the horizontal force dominates the elastic response, even at $\nu=0.45$ and more so as $\nu$ decreases from $1/2$. With regards to validation of the models, most experiments utilize nearly incompressible materials and, as we have stated, one cannot differentiate between Models I and III in this limit. Experimental measurements of the radial displacement field on compressible substrates should resolve this issue once and for all.

A typical measure of the elastic response due to a partially-wetting liquid is the contact-line displacement, which can usually be measured without sophisticated diagnostics. Another benefit is that the contact-line displacement is a scalar measure of the more complicated elastic field. Figure~\ref{fig:CLdispNR} shows how the contact-line displacement for a neutrally-wetting ($\alpha=90^\circ$) substrate varies with the Poisson ratio $\nu$ and solid elastocapillary number $\Upsilon$ for Models I,III. The information shown here could be used in future experiments to reconcile the appropriate contact-line law, either Model I or III.

%%%%%%%%%%%%%%%%%%%%%%%%%%%%%%%%%%%%%%%%%%%%%%%%%%%%%%%%%%%%%%%%%%%%%%%%%%%%%%%%%
\begin{figure}
\begin{center}
\begin{tabular}{cc}
\large{$u^{\mathrm{CL}}_z\sin\alpha$} & \large{$u^{\mathrm{CL}}_r\sin\alpha$}  \\
\includegraphics[width=0.22\textwidth]{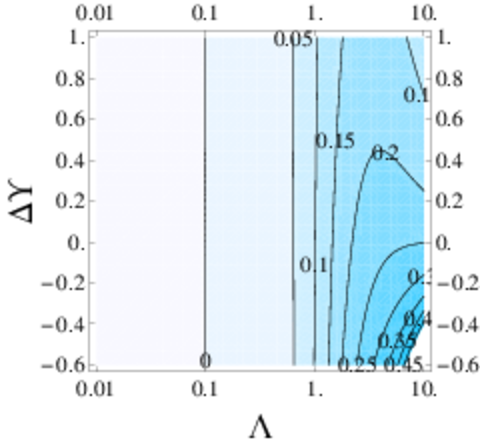} &
\includegraphics[width=0.22\textwidth]{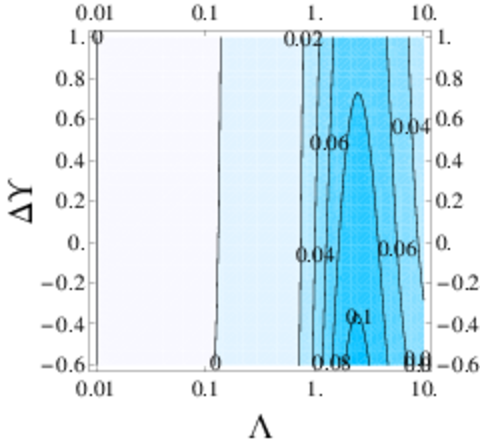}
\end{tabular}
\end{center}
\caption{Comparison of the axial $u^{\mathrm{CL}}_z \sin\alpha$ and radial $u^{\mathrm{CL}}_r \sin\alpha$ contact-line displacement for Models I and III, as it depends upon $\Delta\Upsilon$ and the contact-line radius $\Lambda$ for $\nu=1/2$ and $\Upsilon_{sg}=1$.  \label{fig:CLdisp90DR} }
\end{figure}
%%%%%%%%%%%%%%%%%%%%%%%%%%%%%%%%%%%%%%%%%%%%%%%%%%%%%%%%%%%%%%%%%%%%%%%%%%%%%%%%

Finally, we show how the contact-line displacement varies with $\Lambda$ and $\Delta\Upsilon$ in Figure~\ref{fig:CLdisp90DR}. Note that for partially-wetting situations, both $\alpha$ and $\Delta\Upsilon$ change with the surface chemistry. Hence, we plot the displacements $u\sin\alpha$. The contact-line displacement for Model II is given in the Supplementary Material.  We view Figure~\ref{fig:CLdisp90DR} as a guide for future studies on partially-wetting substrates.

\section{Discussion\label{sec:discussion}}
We have considered the elastic deformations of a soft substrate due to the presence of a partially-wetting liquid. We construct a general solution for the displacement potential (Love function) comparing three rival contact-line models for wetting forces imparted by the liquid onto the solid. In addition, our model generalizes the concept of solid surface tension to partially-wetting substrates $\alpha \neq 90^\circ$, where $\Sigma_{ls}\neq\Sigma_{sg}$. The result of which is a non-standard boundary-value problem that we solve using a dual integral equation. The thrust of this work is that our general solution encompasses all current contact-line models, as well as the interpretation of solid surface tension as a surface stress $\Sigma\equiv\Sigma_{sg}=\Sigma_{ls}$ or surface energy $\Sigma_{sg}\neq\Sigma_{ls}$.

We compare the computed elastic displacement fields to relevant experiments \cite[]{jerison11,style2013universal}, which allows us to identify the most likely model of wetting of soft substrates from the potential candidate models. When comparing to experiment, we immediately see that the surface energy interpretation $\Delta\Upsilon\neq0$ is an essential feature that should be included in any model of partial-wetting. Contact-line Model II, with a horizontal wetting force that neglects the Poisson ratio, is ruled out as a candidate based upon the dramatic differences between the computed displacement field and experimental observations. This leaves contact-line Model I and III with the generalization of solid surface tension $\Delta\Upsilon\neq0$. However, since the relevant experiments involve incompressible substrates $\nu\approx1/2$, which also coincides with the degenerate limit between Models I and III, we are unable to identify the appropriate wetting law at this time. Instead, we use our solution to the general problem to show how
measures of the elastic response vary with the relevant system parameters. The strategy is to use the theory to suggest experimental efforts to resolve this dispute, which is of practical importance in moving the field forward.

\section*{Acknowledgements} The authors are grateful for support from the National Science Foundation under grant
number DMS-0968258.

\appendix
\section{Computation of the constants $C,D$}
The integrals $A_k$, defined in eqn.~(\ref{Akdef}), can be evaluated by interchanging the order of integration with respect to $\mathrm{d}r\mathrm{d}t\rightarrow \mathrm{d}t\mathrm{d}r$ and making use of a Bessel function identity,
\begin{equation} J_0(sr) = \frac{2}{\pi} \int_0^r{\frac{\cos st}{\sqrt{r^2-t^2}}\mathrm{d}t}, \label{Besseldef} \end{equation}
which yields
\begin{eqnarray} && A_1(s) = \Delta \Sigma R \frac{s^6 \sinh s\left(J_0(sR)^2+J_1(sR)^2\right)}{4\left(1-2\nu\right)} \\
&& A_2(s) = \Delta \Sigma \frac{1}{2} R s^5\left(J_0(sR)^2+J_1(sR)^2\right)\times \nonumber \\ && \hspace{1in}\left(s \cosh s + (4\nu-3)\sinh s\right). \end{eqnarray}
Finally, we apply scalings (\ref{scalings}) and simultaneously solve (\ref{HforceBC},\ref{dualsol2}) to give
\begin{eqnarray} C \times X(s)/(2(-1+2\nu))= 2 \hat{F_z}(s) \left(s \cosh s + (-1 + 2 \nu) \sinh s\right) \nonumber\\ + \hat{F_r}(s)\left(2s \sinh s + 4 (-1+\nu)\cosh s +  \Upsilon_{sg}\left(s^2(1+\nu)\cosh s \right. \right. \nonumber\\ \left.\left. + s(-3+\nu+4\nu^2)\sinh s\right)- \Delta \Upsilon \Lambda^2 s (1+\nu)\left(s \cosh s \right.\right.\nonumber \\ \left.+ (-3+4\nu)\sinh s\right)  \left(J_0(s\Lambda)^2+J_1(s\Lambda)^2)\right)\end{eqnarray}

\begin{eqnarray} D \times X(s) = \hat{F_z}(s)\left(2s \sinh s +4(1-\nu)\cosh s\right) \nonumber\\  +\hat{F_r}(s)\left(2s \cosh s  +  (2 - 4 \nu + s^2 \Upsilon_{sg} (1 + \nu) -\right. \nonumber \\ \left. \Delta \Upsilon \Lambda^2 s^2 (1 + \nu) (J_0(s\Lambda)^2+J_1(s\Lambda)^2)) \sinh s\right)\label{CDdef}
\end{eqnarray}
where
\begin{eqnarray}X(s)=s^3 \left(5 + 2 s^2 + 4 \nu (-3 + 2 \nu) + (3 - 4\nu)\cosh 2s  \right.\nonumber \\ \left.+   s \Upsilon_{sg} (-1 + v^2) (2 s + (-3 + 4 v) \sinh 2s) \nonumber \right. \\ \left. -\Delta \Upsilon \Lambda^2 s (-1 + \nu^2) (J_0(s\Lambda)^2+J_1(s\Lambda)^2) \right.\nonumber \\ \left. \times (2 s + (-3 + 4 \nu) \sinh 2s)\right),\label{Xidef} \end{eqnarray}
and $\Delta\Upsilon\equiv \Upsilon_{sg}-\Upsilon_{ls}$. The applied forces $\hat{F}$ are given by
\begin{equation} \hat{F_z}(s) = \sin \alpha \left(\Lambda J_0(s\Lambda)- \frac{2}{s}J_1(s\Lambda)\right),\: \hat{F_r}(s)=F_{\mathrm{CL},r}\Lambda J_1(s\Lambda) , \label{forces}\end{equation}
with the coefficient $F_{\mathrm{CL},r}$ taken from the models given in Table~\ref{tab:Hforce}.

%\footnotesize{
%\bibliography{ElastocapillaryDeformations} %your .bib file
%\bibliographystyle{rsc} %the RSC's .bst file
%}

\providecommand*{\mcitethebibliography}{\thebibliography}
\csname @ifundefined\endcsname{endmcitethebibliography}
{\let\endmcitethebibliography\endthebibliography}{}

\end{document}